\documentclass{article}

\usepackage{algorithm2e}
\usepackage{float}
\usepackage{algpseudocode}
\usepackage{tikz}
\usepackage{amsmath}
\usepackage{pgfplots}
\usepackage{appendix}
\pgfplotsset{compat=1.18}
\usepackage{booktabs} 
\usepackage{adjustbox}
\usepackage{pgfplots}
\usepackage{textcomp}
\usepackage{url}

\pgfplotsset{compat=1.17} 

\newcommand{\remove}[1]{}

\headheight     = \baselineskip
\headsep        = \baselineskip
\textheight     = \paperheight
\textwidth      = \paperwidth
\linewidth      = \textwidth

\begin{document}
\date{}
\title{\Large \bf Sunlight for Darcula: Bypassing Emotet-Based Array Canaries via Autonomous Function Call Resolution}

\author{Nathaniel Oh, Paul Attie, Anas Obeidat\\School of Computer and Cyber Sciences\\Augusta University\\Augusta, Georgia 30912}

\maketitle

\begin{abstract}
We observed the Array Canary, a novel JavaScript anti-analysis technique currently exploited in-the-wild by the Phishing-as-a-Service framework Darcula. The Array Canary appears to be an advanced form of the array shuffling techniques employed by the Emotet JavaScript downloader. In practice, a series of Array Canaries are set within a string array and if modified will cause the program to endlessly loop. In this paper, we demonstrate how an Array Canary works and discuss Autonomous Function Call Resolution (AFCR), which is a method we created to bypass Array Canaries. We also introduce Arphsy, a proof-of-concept for AFCR designed to guide Large Language Models and security researchers in the deobfuscation of "canaried" JavaScript code. We accomplish this by (i) Finding and extracting all Immediately Invoked Function Expressions from a canaried file, (ii) parsing the file's Abstract Syntax Tree for any function that does not implement imported function calls, (iii) identifying the most reassigned variable and its corresponding function body, (iv) calculating the length of the largest string array and uses it to determine the offset values within the canaried file, (v) aggregating all the previously identified functions into a single file, and (vi) appending driver code into the verified file and using it to deobfuscate the canaried file.
\end{abstract}

\section{Introduction}

Phishing\cite{dhamija2006phishing}, or the attempt to trick or coerce a victim into submitted sensitive data to an attacker, is a tale as old as time. However, the approach and techniques used for phishing have adapted and evolved as the internet grew in size and complexity\cite{alkhalil2021phishing}. According to IBM\cite{ibmPhish}, phishing is responsible for 16\% of all data breaches reported in 2024. Nowadays, it's quite common to receive phishing attempts via SMS, email, and social media\cite{chiew2018survey}. 

Recently, security researchers have found that threat actors, who employ phishing attacks at-scale, have compiled a series of phishing tools into what's known as a Phishing Toolkit\cite{sonowal2022phishing}. Phishing Toolkits such as Gophish\cite{githubGitHubGophishgophish} are a collection of tools such as fake login pages, email templates, and much more. 

However, until recently, more sophisticated toolkits that are highly tailored to specific targets were unavailable to lower-resourced cyber-criminals. However, subscription-based Phishing-as-a-Service (Phaas) frameworks have lowered the barrier of entry for access to sophisticated phishing toolkits\cite{opswatPhish}. Now, lower-resourced threat actors have the ability to deploy customized phishing campaigns that no longer have a simplistic and rudimentary approach. 

In particular, a recent PhaaS framework has emerged called Darcula\cite{netcraftShadowsdarcula}. Darcula is a sophisticated PhaaS framework and supports around 200 phishing templates in over 100 different countries with over 20,000 Darcula-affiliated domains\cite{netcraftShadowsdarcula}. Two notable advanced anti-analysis techniques employed by Darcula are JavaScript Array Canaries and User-Agent filtering. While we will primarily discuss the novel Array Canary anti-analysis technique, it is still notable to mention that if a user clicks on a link to a Darcula PhaaS endpoint, Darcula will return a different result if the User-Agent string is not equal to an iPhone User-Agent string similar to \texttt{User-Agent: Mozilla/5.0 (iPhone; U; CPU iPhone OS 4\_3\_2 like Mac OS X; nl-nl) AppleWebKit/533.17.9 (KHTML, like Gecko) Version/5.0.2 Mobile/8H7 Safari/6533.18.5}. 

The first advanced anti-analysis technique, Array Canaries, is a novel implementation of the stack canary first implemented by StackGuard\cite{cowan1998stackguard} in 1998. Back then, a "canary word" was used to detect if a memory stack was corrupted\cite{cowan1999protecting}. However, as time progressed, the canary word, or stack canary, was phased out with the implementation of the shadow stack\cite{sinnadurai2008transparent} due to the increased performance cost of the stack canary\cite{dang2015performance}. However, the idea of using the integrity of a value in a series of values to detect corruption has continued to hold value and merit. Nowadays, instead of memory-based canaries to prevent memory corruption, we observed the implementation of array-based canaries to prevent deobfuscation in sophisticated Phishing-as-a-Service (PhaaS) campaigns.

Cryptographic obfuscation\cite{hofheinz2007obfuscation}, or code obfuscation, is a class of techniques used to obscure and dilute a program's purpose through varying methods including purposeless code, encoded variables, and hard-to-read formatting\cite{behera2015different}. There are two primary categories for code obfuscation: Compile-time language obfuscation\cite{schrittwieser2014covert} and source-code obfuscation\cite{ceccato2014family}. In this paper we will be discussing a specific JavaScript source-code obfuscation technique we observed and how to circumvent it. Therefore, the ability to deobfuscate and understand the contents of a deployed PhaaS framework has become a paramount focus for defenders detecting and preventing such attacks from occurring\cite{reasonlabsWhatDeobfuscation}.

In the past few years, security researchers have observed that JavaScript obfuscation has an increasing presence within phishing campaigns\cite{darkreadingJavaScriptObfuscation}. While obfuscation itself does not imply maliciousness\cite{xu2012power}, obfuscated malicious JavaScript does raise the bar for analysis and reverse engineering efforts. An immense amount of research has been conducted on various methods of dynamic and static JavaScript deobfuscation\cite{6258292, Lee2012SAFEFS, 10.1145/3098954.3107009, githubFakeium, githubExtractJS, JStillPaper, JSHintSpringer, 7163054, 6258292} and malicious JavaScript detection\cite{10.1145/2638404.2737181, NDICHU2019105721, alazab2022detection, ren2023empirical}. 

A specific example of an anti-analysis technique deployed by an obfuscated JavaScript file would be Darcula's use of Array Canaries. At a high level, an array canary takes the following form: \texttt{['1', '2', '4111AjzqZi', '3']}. If the \texttt{4111AjzqZi} value is at any way modified or overwritten, the entire program will endlessly loop and will not deobfuscate itself. However, if the Array Canary values are not changed, the program will deobfuscate itself and run.

Current state-of-the-art machine learning deobfuscation techniques for obfuscated JavaScript code examines the malicious code in its entirety and does not pre-process the submitted sample\cite{ISHIDA2023110138}. This may cause issues as the Array Canary's value cannot be modified and the offset from which the Array Canary calculates itself is determined through an Immediately Invoked Function Expression (IIFE) which isn't immediately apparent when reversing a canaried JavaScript file. 

It has been shown that context-aware grounding of LLMs have led to increased model performance and accuracy\cite{talukdar2024improving}. For example, it has been shown that unaided LLM comprehension struggles with nuanced linguistic features\cite{zhu-etal-2024-large}. As such, there exists a need for recursive prompting that guides an LLM through contextualized guardrails to prevent hallucination and deviation. Recently, advancements have been made in recursive introspection to provide LLM agents the ability to self-correct their own mistakes\cite{qu2024recursive}. Specifically, decomposition prompting has outperformed tradition prompting through sub-task evaluation\cite{khot2022decomposed}. 

Classical symbolic execution engines and abstract analysis techniques have provided extensive benefit and viability in JavaScript deobfuscation efforts\cite{Lee2012SAFEFS, 6258292, 10.1145/2638404.2737181, 10.1145/3098954.3107009, 10.1145/3098954.3107009, githubFakeium, githubExtractJS, JStillPaper, JSHintSpringer, 7163054, 6258292}. However, a major drawback of classical symbolic execution is its inability to generate inputs when the symbolic path constraints along an execution path include formulas that cannot be solved efficiently by a constraint solver\cite{10.1145/2408776.2408795}. When the Array Canary is modified, the program does not terminate and indefinitely loops. While we are not seeking to solve it, it is important to note that we are presented with the unsolvability of the Halting Problem\cite{boyer1984mechanical} due to the infinite looping of a modified Array Canary.

Therefore, to combat Array Canaries and their related IIFE implementation, we propose the Autonomous Function Call Resolution (AFCR) algorithm which can resolve a canaried file by Living-Off-the-Land\cite{satyanarayanancase} and reusing the obfuscated JavaScript code to resolve its own function calls. This, in-turn, bypasses Array Canaries and allows Large Language Models (LLMs) and security researchers the ability to introspect deeper into the de-canaried JavaScript file. 
 
The practice of reusing code is not by any means new or revolutionary\cite{haefliger2008code} as it can lead to cost savings and increased productivity\cite{feitosa2020code}. Recent studies show that malware frequently reuses both techniques and code from other malicious software\cite{7888735, 10430055, 9700530, 9720874, 10427663, 9284089}. This proliferation of shared components and techniques allows for a complementary approach to malware deobfuscation and analysis. For example, anti-analysis techniques were bypassed in the BazarLoader malware family through the reuse and adaptation of its own code to defeat obfuscated API calls\cite{paloaltonetworksDefeatingBazarLoader}.

Given frequent reuse of code in malware, we have reasoning to believe that the proliferation the novel Array Canary technique is imminent if not already occurring. Thus, to counter the spread of Array Canaries, we propose Arphsy\footnote{\url{https://github.com/calysteon/Arphsy}}, an Abstract Syntax Tree parser and aggregator that uses similar techniques to DONAPI\cite{huang2024donapi}, but instead of detecting Malicious NPM packages through Behavior Sequence Knowledge Mapping via API sequences, Arphsy leverages the most reassigned variable in a file to deobfuscate and resolve function calls. 

\begin{table*}[ht!]
\centering
\caption{\texttt{uspass} C2 Deployment Artifacts}
\label{table:uspass}
\begin{tabular}{p{1cm}cccccc}
\hline
\# & Filename & Family & Affiliation & Hosted Domains \\
\hline
1 & \texttt{3c4c0830rSGF6.js} & \texttt{rSGF6} & \texttt{uspass} & \texttt{507} \\
2 & \texttt{4fa45dd1rSGF6.js} & \texttt{rSGF6} & \texttt{uspass} & \texttt{0} \\
3 & \texttt{7d1ef71erSGF6.js} & \texttt{rSGF6} & \texttt{uspass} & \texttt{510} \\
4 & \texttt{09aa6908rSGF6.js} & \texttt{rSGF6} & \texttt{uspass} & \texttt{0} \\
5 & \texttt{09bf01f8rSGF6.js} & \texttt{rSGF6} & \texttt{uspass} & \texttt{509} \\
6 & \texttt{21f80bafrSGF6.js} & \texttt{rSGF6} & \texttt{uspass} & \texttt{0} \\
7 & \texttt{40efa34arSGF6.js} & \texttt{rSGF6} & \texttt{uspass} & \texttt{509} \\
8 & \texttt{57387ccfrSGF6.js} & \texttt{rSGF6} & \texttt{uspass} & \texttt{0} \\
9 & \texttt{2083659erSGF6.js} & \texttt{rSGF6} & \texttt{uspass} & \texttt{0} \\
10 & \texttt{a2f7ed5brSGF6.js} & \texttt{rSGF6} & \texttt{uspass} & \texttt{509} \\
11 & \texttt{a829703arSGF6.js} & \texttt{rSGF6} & \texttt{uspass} & \texttt{0} \\
12 & \texttt{aff47020rSGF6.js} & \texttt{rSGF6} & \texttt{uspass} & \texttt{508} \\
13 & \texttt{b6d96967rSGF6.js} & \texttt{rSGF6} & \texttt{uspass} & \texttt{508} \\
14 & \texttt{b4106d91rSGF6.js} & \texttt{rSGF6} & \texttt{uspass} & \texttt{0} \\
15 & \texttt{c27b6911rSGF6.js} & \texttt{rSGF6} & \texttt{uspass} & \texttt{509} \\
16 & \texttt{db455d90rSGF6.js} & \texttt{rSGF6} & \texttt{uspass} & \texttt{509} \\
17 & \texttt{edff4021rSGF6.js} & \texttt{rSGF6} & \texttt{uspass} & \texttt{510} \\
18 & \texttt{f75bc729rSGF6.js} & \texttt{rSGF6} & \texttt{uspass} & \texttt{0} \\
\end{tabular}
\end{table*}

\section{Background}

\subsection{Darcula In-the-Wild}

\begin{figure}
    \centering
    \includegraphics[width=\linewidth]{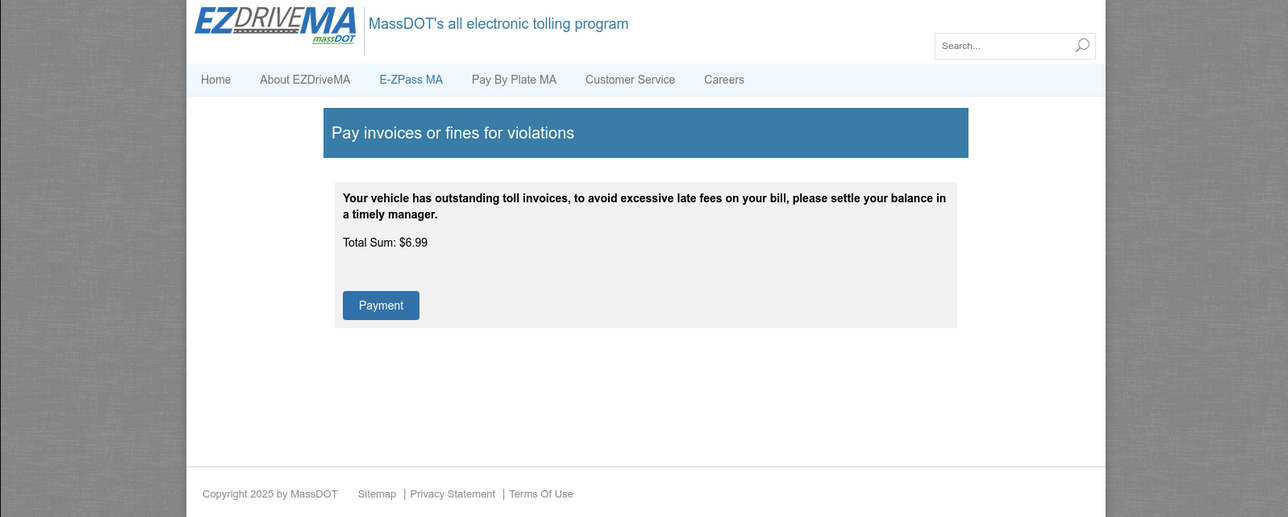}
    \caption{Darcula's Fake Massachusetts Registry of Motor Vehicles Webpage}
    \label{fig:darculaToll}
\end{figure}

A Darcula phishing campaign typically begins when the threat actor sends out an innocuous text message to a victim. If a victim opens the URL, several things may occur. If the browser does not use an iPhone User-Agent string similar to:

\begin{verbatim}
User-Agent: Mozilla/5.0 (iPhone; U; 
CPU iPhone OS 4\_3\_2 like Mac OS X; 
nl-nl) AppleWebKit/533.17.9 (KHTML, 
like Gecko) Version/5.0.2 Mobile/8H7 
Safari/6533.18.5
\end{verbatim}

The server will either return a \texttt{404 error} message or will redirect to a benign webpage such as \texttt{google.com}. Assuming a user is using the iPhone User-Agent string, the user will be redirected to a fake webpage that can imitate government entities such as the United States Postal Service or a local Department of Motor Vehicles. A bogus charge may be listed that the user needs to pay, such as the toll bill for \$6.99 shown in Figure \ref{fig:darculaToll}. At this point, the attacker will perform a Man-in-the-Middle\cite{meyer2004man} attack and attempt to steal any payment information submitted through the fake website. Once the information is stolen from the victim, the attacker will redirect the victim to the valid government website as a means to not alert the victim. 

One of the more interesting Darcula domains we've observed uses the domain \texttt{shopife}. To gain insight into the activity and behavior of \texttt{shopife}, we ran a query using UrlScan.io\cite{urlscan} examining all unique domains that included the filename \texttt{66bcd94a3aHKX.js}. We found that over 8,000 domains requested this file. Searching through the results, we then noticed that every single domain we checked called back to \texttt{shopife} for both the cJS and then initiated a real-time and bi-directional communication stream between the requesting domain and \texttt{shopife} using the \texttt{socket.io} real-time framework. Now, given this information, it's likely that \texttt{shopife} behaves as some sort of Command-and-Control (C2) server for the 8,000 domains that are requesting the \texttt{66bcd94a3aHKX.js} file. Thus, in summary, we've noticed two classes of domains within the deployed Darcula framework: A Command and Control (C2) server and a endpoint. Examining the list of the 8,000 endpoints interacting with the \texttt{shopife} C2, most of them seem to be eCommerce-related websites. For example, the first result, \texttt{pncldcouture} is a clothing shop using the WooCommerce\cite{woocommerce} plugin and the WordPress Content Management System\cite{wordpress}. From an initial analysis, the content hosted on the site itself is seemingly benign and originates from reputable sources. However, the one exception would be the canaried JavaScript (cJS) loaded directly from the \texttt{shopife} C2. Interestingly enough, when we examined the UrlScan.io scan of \texttt{shopife} C2, the scan was unsuccessful and did not return any content aside from an error message and a redirect chain to a \texttt{phish-bypass} endpoint. Since the \texttt{shopife} C2 appears to be affiliated with over 8,000 endpoints, it's possible that it uses some sort of whitelist as a means of anti-debugging. Since we were not the ones to scan the \texttt{shopife} C2, it is possible it employed a similar technique as both the \texttt{ezdrivema} and \texttt{uspass} variants required the User-Agent \texttt{User-Agent: Mozilla/5.0 (iPhone; U; CPU iPhone OS 4\_3\_2 like Mac OS X; nl-nl) AppleWebKit/533.17.9 (KHTML, like Gecko) Version/5.0.2 Mobile/8H7 Safari/6533.18.5} in order to interact with the domain.

\begin{figure}
    \centering
    \includegraphics[width=0.5\linewidth]{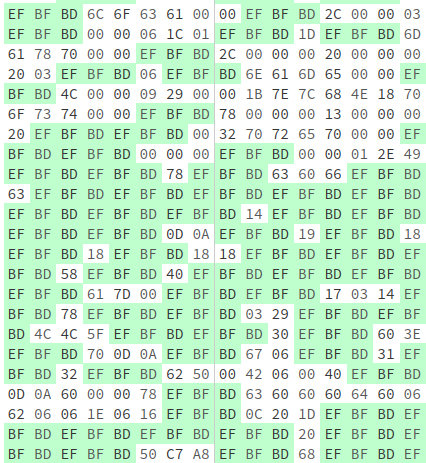}
    \caption{Invalid Unicode Encoding in Darcula WOFF Files}
    \label{fig:efbfbd}
\end{figure}

\begin{figure}
    \centering
    \includegraphics[width=0.75\linewidth]{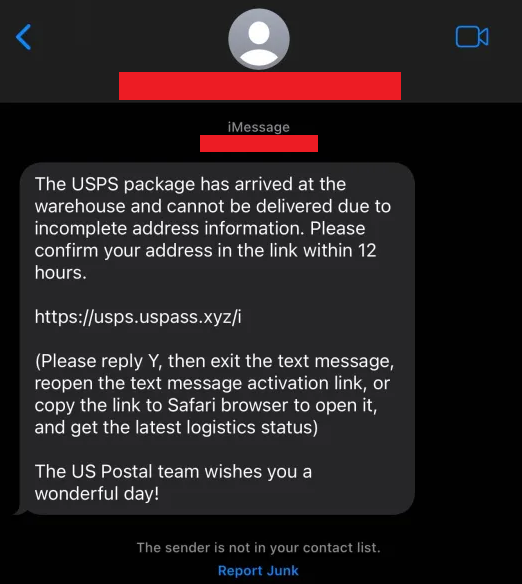}
    \caption{Darcula USPS Phishing Text}
    \label{fig:uspass}
\end{figure}

We've identified 3 separate deployments of the Darcula Phishing-as-a-Serivce Framework. In June 2024, the \texttt{uspass} Darcula variant masqueraded as the United States Postal Service and sent out bogus text messages claiming incomplete address information and thus requested the user to click a link to fill out their information as shown in Figure \ref{fig:uspass}. Unlike the \texttt{ezdrivema} or \texttt{shopife} deployments, \texttt{uspass} was not detected anywhere else and only sought after personal information rather than payment information. While we were unable to record the execution order, the list of files we recovered from the \texttt{uspass} deployment is shown in Table \ref{table:uspass}. Interestingly enough, unlike the centrally-managed \texttt{shopife} C2, the \texttt{uspass} variant seems to self-host all cJS. Thus, \texttt{uspass} take a more decentralized approach and does not rely on a specific C2 to deliver the cJS payload. In addition, the \texttt{uspass} variant is the only Darcula deployment we've noticed that delivers WOFF font files as part of its payload. However, it's unclear how useful the WOFF font files are as they have significant loss of entropy due to invalid Unicode encoding as shown by the \texttt{0xEFBFBD} hex patterns in Figure \ref{fig:efbfbd}.

Each of the identified Darcula variants follows a naming convention that we used to group each family:

\begin{itemize}
    \item \texttt{uspass}: All cJS files end with \texttt{rSGF6}
    \item \texttt{ezdrivema}: All cJS files end with \texttt{pWykA}
    \item \texttt{shopife}: All cJS files end with \texttt{aHKX}
\end{itemize}

Given these subtype identifications, we've mapped the each cJS file in Table \ref{table:ezdrive} to its family, Darcula deployment affiliation, how many domains reference the cJS file, if the file was executed by a visitor, and if so, the order in which it was executed. 

\begin{table*}[ht!]
\centering
\caption{\texttt{shopife} and \texttt{ezdrivema} C2 Deployment Artifacts}
\label{table:ezdrive}
\begin{tabular}{p{1cm}cccccc}
\hline
\# & Filename & Family & Affiliation & Hosted Domains & Executed & Order\\
\hline
1 & \texttt{8134d22dpWykA.js} & \texttt{pWykA} & \texttt{ezdrivema} & \texttt{4} & \texttt{TRUE} & \texttt{1} \\
2 & \texttt{a980d14dpWykA.js} & \texttt{pWykA} & \texttt{ezdrivema} & \texttt{4} & \texttt{TRUE} & \texttt{2} \\
3 & \texttt{736707ecpWykA.js} & \texttt{pWykA} & \texttt{ezdrivema} & \texttt{4} & \texttt{TRUE} & \texttt{3} \\
4 & \texttt{88b36d32pWykA.js} & \texttt{pWykA} & \texttt{ezdrivema} & \texttt{4} & \texttt{TRUE} & \texttt{4} \\
5 & \texttt{09bf01f8pWykA.js} & \texttt{pWykA} & \texttt{ezdrivema} & \texttt{4} & \texttt{TRUE} & \texttt{5} \\
6 & \texttt{edff4021pWykA.js} & \texttt{pWykA} & \texttt{ezdrivema} & \texttt{4} & \texttt{TRUE} & \texttt{6} \\
7 & \texttt{e134e940pWykA.js} & \texttt{pWykA} & \texttt{ezdrivema} & \texttt{4} & \texttt{TRUE} & \texttt{7} \\
8 & \texttt{997b1b1dpWykA.js} & \texttt{pWykA} & \texttt{ezdrivema} & \texttt{4} & \texttt{TRUE} & \texttt{8} \\
9 & \texttt{c27b6911pWykA.js} & \texttt{pWykA} & \texttt{ezdrivema} & \texttt{4} & \texttt{TRUE} & \texttt{9} \\
10 & \texttt{37f0e597pWykA.js} & \texttt{pWykA} & \texttt{ezdrivema} & \texttt{4} & \texttt{TRUE} & \texttt{10} \\
11 & \texttt{35586d99pWykA.js} & \texttt{pWykA} & \texttt{ezdrivema} & \texttt{4} & \texttt{FALSE} &\\
12 & \texttt{4c4bf7ebpWykA.js} & \texttt{pWykA} & \texttt{ezdrivema} & \texttt{4} & \texttt{FALSE} &\\
13 & \texttt{70239002pWykA.js} & \texttt{pWykA} & \texttt{ezdrivema} & \texttt{4} & \texttt{FALSE} &\\
14 & \texttt{a8f9f8efpWykA.js} & \texttt{pWykA} & \texttt{ezdrivema} & \texttt{4} & \texttt{FALSE} &\\
15 & \texttt{8f6a41bdpWykA.js} & \texttt{pWykA} & \texttt{ezdrivema} & \texttt{4} & \texttt{FALSE} &\\
16 & \texttt{6342f2e8pWykA.js} & \texttt{pWykA} & \texttt{ezdrivema} & \texttt{4} & \texttt{FALSE} &\\
17 & \texttt{e1d85e8epWykA.js} & \texttt{pWykA} & \texttt{ezdrivema} & \texttt{4} & \texttt{FALSE} &\\
18 & \texttt{86cd325bpWykA.js} & \texttt{pWykA} & \texttt{ezdrivema} & \texttt{4} & \texttt{FALSE} &\\
19 & \texttt{67eea806pWykA.js} & \texttt{pWykA} & \texttt{ezdrivema} &\texttt{4} &  \texttt{FALSE} &\\
20 & \texttt{loadDarcula.js} & \texttt{N/A} & \texttt{shopife} & \texttt{8881} & \texttt{TRUE} & \texttt{1} \\
21 & \texttt{darcula.js} & \texttt{N/A} & \texttt{shopife} & \texttt{8751} & \texttt{TRUE} & \texttt{2} \\
22 & \texttt{66bcd94a3aHKX.js} & \texttt{aHKX} & \texttt{shopife} & \texttt{8457} & \texttt{TRUE} & \texttt{3} \\
23 & \texttt{83a262943aHKX.js} & \texttt{aHKX} & \texttt{shopife} & \texttt{7934} & \texttt{TRUE} & \texttt{4} \\
24 & \texttt{5c0f42a33aHKX.js} & \texttt{aHKX} & \texttt{shopife} & \texttt{7681} & \texttt{TRUE} & \texttt{5} \\
25 & \texttt{a07e1e913aHKX.js} & \texttt{aHKX} & \texttt{shopife} & \texttt{7431} & \texttt{TRUE} & \texttt{6} \\
26 & \texttt{97c0066d3aHKX.js} & \texttt{aHKX} & \texttt{shopife} & \texttt{7431} & \texttt{TRUE} & \texttt{7} \\
27 & \texttt{1cb699f63aHKX.js} & \texttt{aHKX} & \texttt{shopife} & \texttt{7187} & \texttt{TRUE} & \texttt{8} \\
28 & \texttt{309f99503aHKX.js} & \texttt{aHKX} & \texttt{shopife} & \texttt{6519} & \texttt{TRUE} & \texttt{9} \\
29 & \texttt{13569cd53aHKX.js} & \texttt{aHKX} & \texttt{shopife} & \texttt{6021} & \texttt{TRUE} & \texttt{10} \\
\end{tabular}
\end{table*}

\subsection{Emotet's Influence}

First identified in 2014\cite{trendmicroEMOTETThreat}, Emotet was originally designed as a banking trojan but soon evolved into a downloader for other malware\cite{cisaEmotetMalware}. Known for its sophistication, this polymorphic malware primarily spreads through malicious email attachments and embedded links\cite{kuraku2020emotet}. 

Similarly, Darcula's propagation method is also through embedding links sent in RCS and iMessages\cite{netcraftShadowsdarcula}. Before Darcula was tracked by NetCraft in 2024\cite{netcraftShadowsdarcula}, Emotet's approach to JavaScript obfuscation heavily resembled that of the open-source GitHub project JavaScript Obfuscator\cite{githubGitHubJavascriptobfuscatorjavascriptobfuscator}. However, while Darcula does employ similar techniques, the addition of the Array Canary significantly differentiates it from any publicly-available open-source GitHub project. 

However, the key similarity between both Emotet and Darcula lies within how both frameworks use IIFEs. Fundamentally, the Array Canaries found in Darcula are simply an advanced implementation of the array shuffling technique used in the Emotet JavaScript downloader as shown in Listing \ref{emotet}:

\begin{verbatim}
(function(c, d) {
    var e = function(f) {
        while (--f) {
            c['push'] (c['shift']());
        }
    };
    e(++d);
} (a, 0xea));
\end{verbatim}

The Emotet JavaScript downloader's anti-analysis techniques were primarily targeting human-in-the-loop debugging methods such as in-browser debugging and console logging\cite{maxkerstenEmotetJavaScript}. However, machine learning is becoming a common approach to detecting and deobfuscating JavaScript malware\cite{ndichu2020deobfuscation, sohan2020systematic, dabral2017malicious, 10.1145/2714576.2714620, 10.1145/2666652.2666657, 10.1145/2381896.2381901}. Thus, the approach shifted from simply preventing console logs and general debug actions to integrity checking the array shuffle to ensure that it has not been tamper with before the program executes its payload. 

\subsection{Array Canaries in Darcula}

Similar to the Emotet JavaScript downloader, Darcula uses an IIFE to shuffle an array of strings prior to payload execution. However, it goes one step further and does not simply decrement a counter, but rather loops infinitely until a specific value is reached through a series of \texttt{parseInt} operations. These operations as shown in Listing \ref{acIIFE}, act as a lock where each array shuffle is a key. If none of the Array Canaries or modified, the key will eventually be calculated and the payload will be executed. However, if even one of the Array Canaries change, the entire program loops infinitely. 

\begin{verbatim}
(function (u, A) {
    const h = a0A,
        E = u();
    while (!![]) {
        try {
            const D =
                (parseInt(h(0x154)) / 0x1) * 
                (-parseInt(h(0x152)) / 0x2) +
                (-parseInt(h(0x156)) / 0x3) * 
                (parseInt(h(0x162)) / 0x4) +
                -parseInt(h(0x15b)) / 0x5 +
                -parseInt(h(0x151)) / 0x6 +
                parseInt(h(0x15e)) / 0x7 +
                (parseInt(h(0x159)) / 0x8) * 
                (parseInt(h(0x157)) / 0x9) +
                (parseInt(h(0x15f)) / 0xa) * 
                (parseInt(h(0x160)) / 0xb);
            if (D === A) break;
            else E["push"](E["shift"]());
        } catch (v) {
            E["push"](E["shift"]());
        }
    }
})(a0u, 0x6f0ff);
\end{verbatim}

Now, let's examine how an Array Canary works in Listing \ref{AC1} by simplifying its implementation and resolving all programmatic steps up until the array shifting cycle starts:

\begin{verbatim}
(function (u, A) {
    const h = [ "763343ZEEmqI", "10MjwbHE", "9850357GcgXRv", 
    "VALUE_7", "143668KLzuHC", "2744166fvKFHm", "159958nePvPH", 
    "VALUE_1", "1UiidKZ", "VALUE_2", "51QZYsCO", "9rBvnZg", 
    "VALUE_3", "6312632cMablh", "VALUE_4", "953875DutVaJ", 
    "VALUE_5", "VALUE_6", ];

    while (true) {
        try {
            const D =
                (parseInt(h[3]) / 1) * 
                (-parseInt(h[1]) / 2) +
                (-parseInt(h[5]) / 3) * 
                (parseInt(h[17]) / 4) +
                -parseInt(h[10]) / 5 +
                -parseInt(h[0]) / 6 +
                parseInt(h[13]) / 7 +
                (parseInt(h[8]) / 8) * 
                (parseInt(h[6]) / 9) +
                (parseInt(h[14]) / 10) * 
                (parseInt(h[15]) / 11);
            if (D === 0x6f0ff) break;
            else h["push"](h["shift"]());
        } catch (v) {
            h["push"](h["shift"]());
        }
    }
})(a0u, 0x6f0ff);
\end{verbatim}

We see that a series of operations occur until the total value of \texttt{D} is equal to \texttt{0x6f0ff}. Without shifting the array, let's see what happens if \texttt{D} is not equal to \texttt{0x6f0ff} as shown in Listing \ref{AC2}:

\begin{verbatim}
(NaN / 1) * (-10 / 2) + (-2744166 / 3) * 
(NaN / 4) + -51 / 5 + -763343 / 6 + 6312632 
/ 7 + (1 / 8) * (159958 / 9) + (NaN / 10) * 
(953875 / 11);
\end{verbatim}

Any value that does not precede with a number is set as \texttt{NaN} whereas any value such as \texttt{763343ZEEmqI} is considered a number and any non-numeric values are truncated. Thus, the array will continually shuffle until it is in the correct position where the total value is equal to \texttt{0x6f0ff}. While this appears straight-forward after the fact, the presence of an Array canary throws off the proper placement of each of the array's strings within the canaried file. Therefore, to accurately deobfuscate a file that's been canaried, it's essential to determine the right string array order first. 

\section{Arphsy: Design and Methodology}

\begin{figure*}
    \centering
    \includegraphics[width=0.85\textwidth]{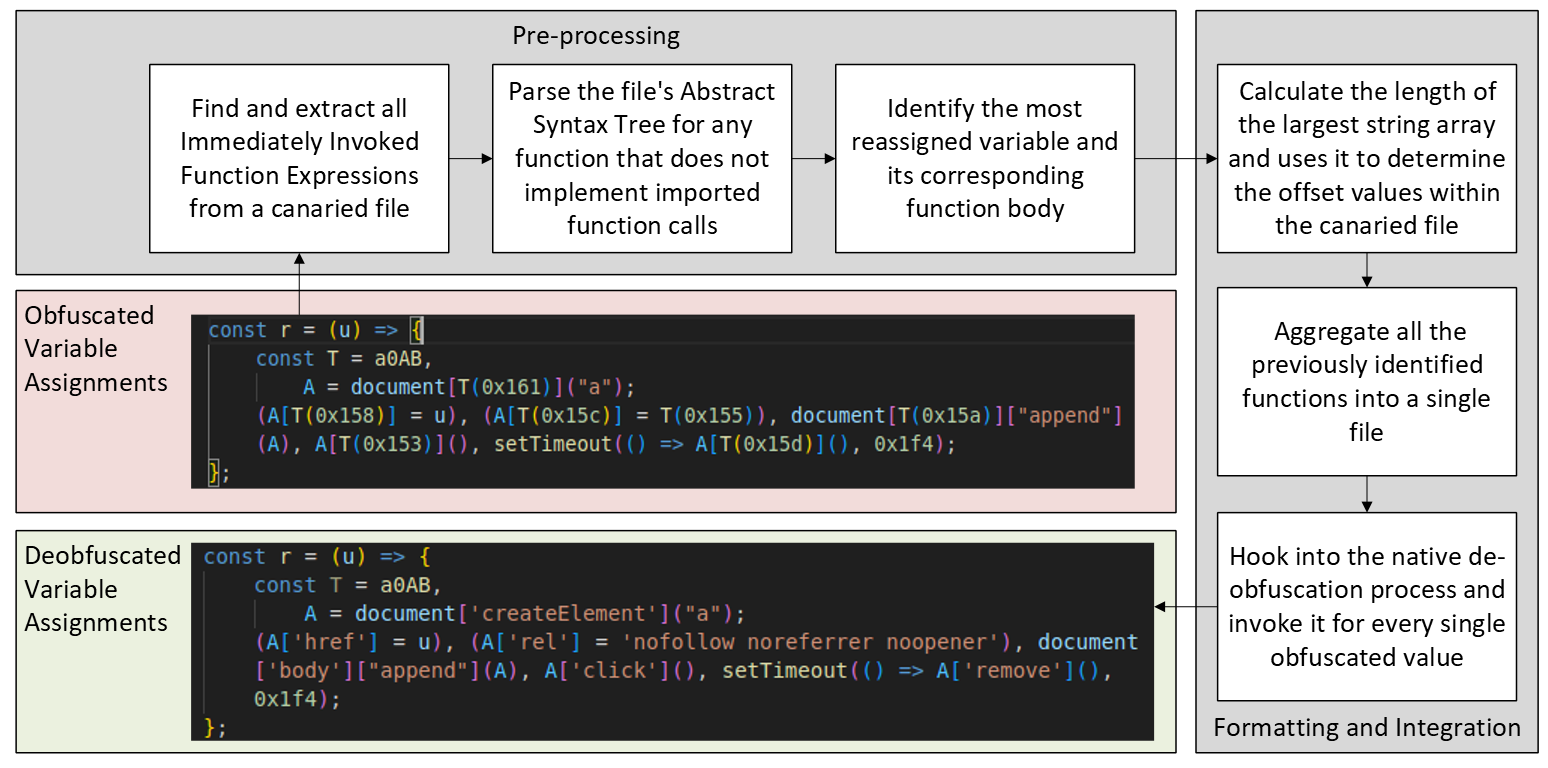}
    \caption{An Overview of Arphsy}
    \label{fig:arphsyOverview}
\end{figure*}

Arphsy is a staged implementation of our proposed AFCR algorithm. Therefore, we will discuss in this section the design behind Arphsy by walking through how AFCR mitigates the anti-analysis behavior of Array Canaries as shown in Figure \ref{fig:arphsyOverview}. At a high level, the Autonomous Function Call Resolution (AFCR) algorithm is designed to systematically bypass JavaScript Array Canaries by leveraging the inherent structure of canaried files. The first step in AFCR involves identifying and extracting all Immediately Invoked Function Expressions (IIFEs) present in the target file. Then, after the IIFEs are extracted, AFCR then parses the target file's Abstract Syntax Tree (AST), analyzing its structure. Within the parsed AST, AFCR searches for any functions that do not invoke imported function calls. Next, AFCR finds the most reassigned variable and resolves its function definition. The most reassigned variable's function definition contains the entry-point to the decanarying routine as well as the initial hexadecimal offset used to calculate which values are to be decanaried. The second hexadecimal offset is found by first calculating the length of the largest array in the canaried file and then obtaining the difference from the first offset. Finally, the extracted IIFE, most reassigned variable's function definition, and the function containing the largest string array is added into a harness file with driver code appended to it. This harness file is then invoked on each value within the calculated hexadecimal offset range to resolve each of the canaried function calls. 

\subsection{Find and Extract IIFEs}

Our first step in AFCR is to identify and extract all IIFEs present in a target file as shown in Listing \ref{findIIFEs}. To accomplish this, we define a function called \texttt{findIIFEs} which parses the contents of the target file into a string variable called \texttt{jsCode}. We then map the variable into an AST using the Acorn library. We recursively navigate through each node of the AST, examining if it is a \texttt{CallExpression} and a \texttt{FunctionExpression} or a \texttt{ArrowFunctionExpression}. If so, we capture the IIFE as the variable \texttt{iifeCode} and push it to the array \texttt{iifes}. Once all child nodes have been traversed, the function returns the array stored in \texttt{iifes}. 

\begin{verbatim}
//...
function findIIFEs(filePath) {
//...
function traverse(node) {
  if (node.type === "CallExpression" && 
    (node.callee.type === "FunctionExpression" || 
    node.callee.type === "ArrowFunctionExpression")) {
        const before = jsCode[node.start - 1] === 
            "(" ? node.start - 1 : node.start;
        const after = jsCode[node.end] === ";" 
            ? node.end + 1 : (jsCode[node.end] === ")" ?
            node.end + 1 : node.end);
        const iifeCode = jsCode.slice(before, after).trim();
        iifes.push(iifeCode);
  }
//...
\end{verbatim}

\subsection{Parsing the AST}

Next, we used similar AST parsing techniques as in \texttt{findIIFEs}. In Listing \ref{parseJSFile}, we create a function called \texttt{parseJSFile} which traverses the AST looking for any node of type \texttt{FunctionDeclaration}, \texttt{VariableDeclaration}, or \texttt{VariableDeclarator}. If the matched node is of type \texttt{FunctionDeclaration}, its name is pushed to the \texttt{functions} array. If the matched node is of type \texttt{VariableDeclaration}, any declared variable name is pushed to the \texttt{variables} array. Finally, if the matched node is of type \texttt{VariableDeclarator} and initializes a function or arrow function, its name is pushed to the \texttt{functions} array. Once all child nodes have been traversed, the function returns both the \texttt{functions} and \texttt{variables} array.

\begin{verbatim}
//...
function parseJSFile(filePath) {
//...
function traverse(node) {
switch (node.type) {
    case "FunctionDeclaration": 
        if (node.id) {functions.push(node.id.name);} break;
    
    case "VariableDeclaration": 
        node.declarations.forEach((declaration) => 
            {if (declaration.id && declaration.id.name) 
                {variables.push(declaration.id.name);}}); 
            break;
    
    case "VariableDeclarator": 
        if (node.init && (node.init.type === "FunctionExpression" 
            || node.init.type === "ArrowFunctionExpression")) 
            {functions.push(node.id.name);}break;
}
//...
\end{verbatim}

\subsection{Function Filtering}

Our next step shown in Listing \ref{filterFunctions} is to import the \texttt{parseJSFile} and use it get a list of defined functions and variables. We again parse the target file's AST. However, this time, when we traverse each AST node, we check if it is of type \texttt{FunctionDeclaratoin} and has a body. If so, we check the function body to see if it only uses functions listed in \texttt{parseJSFile}'s output. If so, we add the parsed function to the \texttt{validFunctions} array. After all the child nodes have been traversed, the function returns the \texttt{validFunctions} array.

\begin{verbatim}
//...
function filterFunctions(filePath) {
//...
function isValidIdentifier(name) 
    {return definedFunctions.includes(name) 
    || definedVariables.includes(name);}
function traverse(node) {
  if (node.type === "FunctionDeclaration" && node.body) {
    const functionName = node.id.name;
    const isValid = checkFunctionBody(node.body);
  
    if (isValid) { 
        const functionCode = code.slice(node.start, node.end); 
        validFunctions.push(functionCode);}
    }
//...
function checkFunctionBody(bodyNode) {
  let isValid = true;
  function validateNode(node) {
    if (node.type === "Identifier") {if 
        (!isValidIdentifier(node.name)) {isValid = false;}}
//...
\end{verbatim}

\subsection{Find the Most Reassigned Variable}

To find the most reassigned variable within a file as shown in Listing \ref{mvr}, we defined \texttt{findMostReassignedVariabe} which begins by parsing the AST of the target file. Each node within the target AST is then parsed and examined to see if it is of type \texttt{VariableDeclarator} or \texttt{AssignmentExpression}. If it is of type \texttt{VariableDeclarator}, if the initializer is a variable, we increment a counter for that specific variable within the \texttt{variableReassignmentCounts} array. Next, if the node is an \texttt{AssignmentExpression}, if the right-hand side of the expression is a variable, the counter for the variable is incremented within the \texttt{variableReassignmentCounts} array. Once all the child nodes are recursively parsed, the most reassigned variable is determined by finding the counter with the highest value in the \texttt{variableReassignmentCounts} array. 

\begin{verbatim}
function findMostReassignedVariable(filePath) {
//...
  function countVariableReference(name) {
    if (!variableReassignmentCounts[name]) {variableReassignmentCounts[name] = 0;    }
    variableReassignmentCounts[name]++;
  }

  function traverse(node) {
    if (!node) return;
    switch (node.type) {
      case "VariableDeclarator":
        if (node.init && node.init.type === "Identifier") 
        {countVariableReference(node.init.name);} break;

      case "AssignmentExpression": 
        if (node.right && node.right.type === "Identifier") 
        {countVariableReference(node.right.name);} break;
//...
  const mostReassigned = Object.entries(variableReassignmentCounts).reduce((max, 
    [name, count]) => (count > max.count ? { name, count } 
    : max),{ name: null, count: 0 });
//...
\end{verbatim}

After the most reassigned variable is discovered, it's function definition is then obtained as shown in Listing \ref{mrvAssign}. This is done by against parsing the target file's AST. Each node is first checked to see if it is of type \texttt{VariableDeclarator}, equal to the name of the most reassigned variable, initialized, and either of type \texttt{FunctionExpression} or \texttt{ArrowFunctionExpression}. Next, the examined AST node is checked to see if it is of type \texttt{FunctionDeclaration} and equal to the most reassigned variable's name. If either of these conditions is true, then the source code from that function is saved to the \texttt{functionDefinition} variable. Once all the child nodes have been recursively parsed, the contents of the \texttt{functionDefinition} variable is appended to the harness file alongside a export statement which provides the name of the most reassigned variable.

\begin{verbatim}
function findFunctionDefinition(filePath, varName) {
//...
if (node.type === "VariableDeclarator" && 
    node.id.name === varName && node.init && 
    (node.init.type === "FunctionExpression" || 
    node.init.type === "ArrowFunctionExpression")) 
    {functionDefinition = sourceCode.slice(node.start, node.end);}

if (node.type === "FunctionDeclaration" && 
    node.id.name === varName) 
    {functionDefinition = sourceCode.slice(node.start, node.end);}
//...
\end{verbatim}

\subsection{Harness Development}

Because the canaried file oftentimes uses hex values in other operations aside from Array Canarying, bounds need to be established to be able to accurately determine which hexadecimal values are used in Array Canarying and which ones are used for normal operations by the canaried file. To do this, we will find two hex values: The first is the offset used by the most reassigned variable's function and the second is the length of the largest string array added to the first offset found. Thus, to find the first hex value we will create a function called \texttt{renameFirstExport} which renames the most reassigned variable found in the export statement to \texttt{PLACEHOLDER}. Now, we will then search the PLACEHOLDER function, previously the most reassigned variable's name, for the hex offset used by the Array Canary. 

\begin{figure}[ht!]
    \centering
    \includegraphics[width=0.5\linewidth]{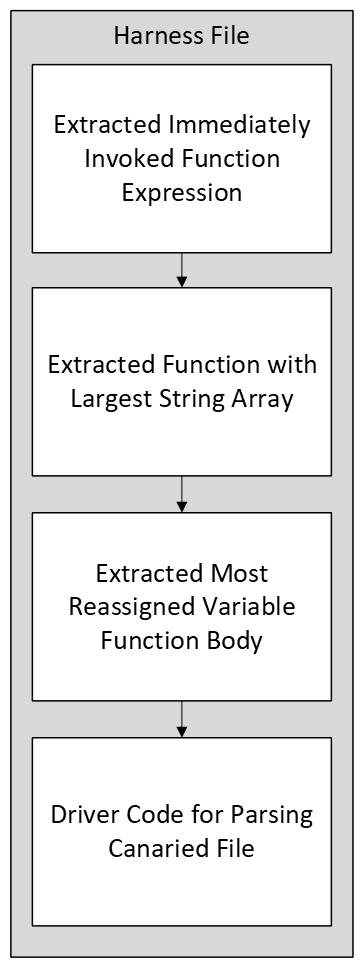}
    \caption{Harness File Structure}
    \label{fig:harnessFile}
\end{figure}

Once the hex offset is found, a function called \texttt{findLargestStringArray} is used to find the largest string array within the canaried file. Once the array is found, the length is returned. This length is added to the first hex offset found to then equal the second hex offset value. Both of these values are then appended to the harness file. Finally, to complete the harness file, driver code is added to find each hex value within the hex range we found previously. The hex value found will then be passed as an argument to the most reassigned variable's function and the return value is the resolved function call. This process is repeated for each hex value within the defined range, which in-turn resolves all obfuscated function calls as shown in Figure \ref{fig:harnessFile}

\section{Discussion and Future Work}

\begin{figure}
    \centering
    \includegraphics[width=0.75\linewidth]{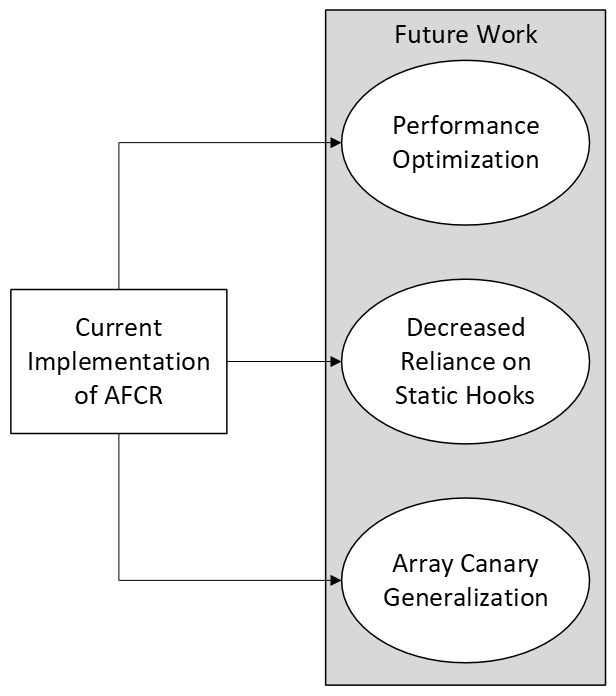}
    \caption{Areas of Future Work}
    \label{fig:futureWork}
\end{figure}

The Darcula framework's use of Array Canaries demonstrates a significant advancement in JavaScript anti-analysis techniques. Through the implementation of AFCR, we show a possible bypass defenders can use when deobfuscating canaried files. However, there still exists limitations in our approach as shown in Figure \ref{fig:futureWork}.

Given the complexity and sophistication of the Darcula PhaaS, there are still many aspects which we have not discussed. Our focus in this paper centered around the implementation of Array Canaries and User-Agent filtering. However, we have observed additional behavior such as unrecognized Secure Sockets Layer messages and legacy JavaScript variants. In addition, we did not fully deobfuscate the more complex files over 1,000 lines long. Thus, it is possible that other techniques or vulnerabilities are exploited that we have not uncovered yet.

Our primary focus when developing AFCR was to identify key sites we can hook into statically to facilitate the dynamic analysis. Thus, certain signatures were found in the Darcula cJS files that, if changed, would easily throw off the current approach to AFCR. For example, when searching for the first hex offset, if the target function is overridden, the AFCR algorithm would either fail to resolve or not resolve the proper hex offset. This may potentially be resolved by leveraging an LLM to find the hex offset that is used by the IIFE callee, but we have not verified if this approach is viable. 

The biggest challenge we have found is in navigating canaried files accurately and quickly. While it theoretically is possible to recreate each step through an LLM prompt, we have found that LLMs tend to simplify complex programs in nonsensical ways. For example, we submitted a query containing parenthesis but the response returned converted the parenthesis to square brackets. Furthermore, the variance in response can pose challenges in properly contextualizing the query. For example, in previous iterations of Arphsy, we relied on the use of LLMs to verify that the harness file we were creating was syntactically correct.  Initially we used the following prompt: 

\begin{verbatim}
Fix this JavaScript file but only respond 
with the fixed file.
\end{verbatim}

However, the LLM kept adding in \texttt{this} statements and changed the variable names. So, we updated the prompt to:

\begin{verbatim}
Fix this JavaScript file but only respond 
with the fixed file. However only complete 
parenthesis or other small formatting issues. 
Add all missing semi-colons, especially after 
an IIFE. Do not change any value or add 
the keyword this
\end{verbatim}

However, this only worked for small files. Development was on a case-by-case basis as we encountered variants of previously identified IIFEs and other components needed, the LLM would respond with different hallucinations and inconsistencies. Thus, due to the changing nature of what the LLM got wrong, we decided to remove it entirely from AFCR and exclusively rely on AST parsing and Regular Expression (RegEx) hooking. 

The downside of our reliance on RegEx is simply that future iterations of Darcula can easily change the hooks RegEx uses and thus prevent the current form of AFCR from completing. Even with this in consideration, due to the massive number of deployed Darcula domains, we still will release Arphsy in its current state with the hopes that defenders and security researchers can use it to better protect their network and the Internet at large. 

Currently, as the AFCR algorithm is restricted by variant-specific hooks, an important area of research would be in generalizing the AFCR approach to encompass more AST-centric analysis to circumvent any reliance on static patterns found within the canaried file. In addition, due to the similarity of the Darcula's Array Canaries with the Emotet JavaScript downloader's array shuffling, it may be possible to back-port AFCR to work on less sophisticated implementations of array-based JavaScript obfuscation techniques. 

LLMs serve an important role in deobfuscating JavaScript\cite{patsakis2024assessing, zaharia2024gview, sridhara2023chatgpt}. However, we did not explore the capabilities of LLM deobfuscation due to frequent program flow changes and syntax inconsistencies. It still stands to reason, that with enough contextualization and framing, it may be possible to use LLMs effectively to bypass previous constraints held by the static RegEx hooking currently used by AFCR

Another avenue for future work could be optimizing performance. Currently, excessive read and write operations occur which at-scale may cause issues given the size and scope of files. Thus, it may be beneficial to look into how file pre-processing could occur by minimizing, chunking, or removing parts of code that is not of interest. For example, if a large file could be broken up into smaller, independent files, that may ease the computational burden of processing a single monolithic file. Furthermore, extensive use of AST parsing is employed and realistically the number of operations can be minimized. Further work may be able to reduce the number of times the AST is parsed and create a cache of sorts that can be internally referenced by the AFCR algorithm when needed to revisit a particular node or data-point. Finally, the last area we will discuss possible performance optimizations would be in parallelizing the AST parsing and related operations. If the boundary between parallelizable and non-parallelizable code can be drawn, then further modifications can be made the how the AFCR algorithm is executed to increase performance.

\section*{Acknowledgments}

We would like to thank both Angelica Reeser and Victor Haugen for their contributions to our research. 

\section*{Ethics considerations}

The Darcula PhaaS framework is an active threat with over 20,000 tracked domains\cite{netcraftShadowsdarcula}. Thus, it is of the upmost importance to triage, assess, and understand effective mitigations and ways to protect both individual networks and the Internet as a whole. Currently, research is limited to simply notifying the public of the phishing risk and the targeted approach to steal sensitive information such as login credentials or financial data\cite{phishfirewallUnderstandingDarcula}. There is not a comprehensive study on the implementation, deployment, or usage of the Darcula-specific anti-analysis techniques such as Array Canaries or how Darcula uses User-Agent filtering to exclusively target iPhones. We have yet to identify the reason why iPhones are the primary target of both the \texttt{uspass} and \texttt{ezdrivema} campaigns. However, it remains paramount to share with the community of defenders and security researchers the information we've already uncovered. We recognize the risk associated with drawing attention to novel anti-analysis techniques used by a threat actor in obfuscating JavaScript. To mitigate this risk, our focus was not only on showing where Array Canaries are currently being used in-the-wild but also to provide a solution through AFCR. While our proposed approach is by no means comprehensive, our focus was to demonstrate a possible Array Canary bypass that can be used by defenders and security researchers. Thus, by releasing the source code to Arphsy, we hope it helps the cybersecurity community as a whole gain deeper introspection into Darcula and similar frameworks that use Array Canaries. 

\section*{Open science}

All code and artifacts needed for proof-of-concept are available online at \url{https://github.com/calysteon/Arphsy}. In addition, to demonstrate the deobfuscation capability of AFCR, we made available a series of sample files that implement the Array Canary obfuscation technique used by the Darcula Phishing-as-a-Service framework.

\bibliographystyle{plain}
\bibliography{main}
\end{document}